\documentclass[aps,prb,onecolumn,showpacs,groupedaddress,amsmath,floatfix]{revtex4}

\usepackage{graphicx}

\begin{document}

\title{Effects of lateral diffusion on morphology and dynamics of a
  microscopic lattice-gas model of pulsed electrodeposition}
\author{Stefan Frank}
\affiliation{Center for Materials Research and Technology}
\affiliation{School of Computational Science}
\affiliation{Department of Physics, Florida State University,
  Tallahassee, FL 32306-4350, USA}

\author{ Daniel E. Roberts}
\affiliation{Center for Materials Research and Technology}
\affiliation{School of Computational Science}
\affiliation{Department of Physics, Florida State University,
  Tallahassee, FL 32306-4350, USA}

\author{Per Arne Rikvold}
\affiliation{Center for Materials Research and Technology}
\affiliation{School of Computational Science}
\affiliation{Department of Physics, Florida State University,
  Tallahassee, FL 32306-4350, USA}
\date{\today}

\begin{abstract}
The influence of nearest-neighbor diffusion on the decay of a
metastable low-coverage phase (monolayer adsorption) in a square
lattice-gas model of electrochemical metal deposition is investigated
by kinetic Monte Carlo simulations. The phase-transformation
dynamics are compared to the well-established
Kolmogorov--Johnson--Mehl--Avrami theory [A.~N. Kolmogorov,
  Bull. Acad. Sci. USSR, Phys. Ser. {\bf 1},  335  (1937);
  W.~A. Johnson and R.~F. Mehl,
  Trans. Am. Inst. Min. Metall. Eng. {\bf 135},  416 (1939);
  M. Avrami, J. Chem. Phys. {\bf 7},  1103  (1939); {\bf 8},  212
  (1940); {\bf 9},  177  (1941)]. The phase
transformation is accelerated by diffusion, but remains in accord with
the theory for continuous nucleation up to moderate diffusion rates. At
very high diffusion rates the phase-transformation kinetic shows
a crossover to instantaneous
nucleation. Then, the probability of medium-sized clusters is reduced in
favor of large clusters. Upon reversal of the supersaturation, the
adsorbate desorbs, but large clusters still tend to grow during the
initial stages of desorption. Calculation of the free energy of
subcritical clusters by enumeration of lattice animals yields a
quasi-equilibrium distribution which is in reasonable agreement
with the simulation results. This is an improvement relative to classical
droplet theory, which fails to describe the distributions, since the
macroscopic surface tension is a bad approximation for small clusters.
\end{abstract}


\maketitle

\section{Introduction}
\label{par:introduction}
Kinetic lattice-gas models are a valuable tool for the modeling of
nucleation and growth of adsorbate
layers.\cite{BrownBook99,RikvoldBook02,Ratsch03} In electrochemical
systems, due to the presence of solvents and counter ions, common
models are often more severe simplifications compared to surface
science in ultra-high vacuum. It is also more difficult to estimate
model parameters from first principles.\cite{Wang02,
  Wang04_unpublished} Though such estimations might seem
desirable, they are not mandatory, and simpler ones, for example by fits to
experiments, may be used, at least as a start.\cite{Mitchell01, Hamad03} In spite of these
complications, successful applications of the method
to electrochemical systems are encouraging~(see, e.g., Refs.~\onlinecite{Koper98,%
  Mitchell01}). In addition, taking a more fundamental
point of view, such truly microscopic models -- and the equivalent two-dimensional Ising models --
may be compared to prevalent theories of nucleation and growth like the
Kolmogorov--Johnson--Mehl--Avrami (KJMA) theory of phase
transformations,\cite{Kolmogorov37, Johnson39, Avrami39, Avrami40,
  Avrami41} thus enabling one to investigate the latter's validity and 
underlying assumptions.\cite{Rikvold94, Ramos99,
  Shneidman99a} The validity of the KJMA theory has been
established not too far below the critical temperature over a wide
range of magnetic fields (corresponding to supersaturation in the
lattice-gas language).\cite{Rikvold94, Ramos99,
  Shneidman99a} However, only spin-flip (adsorption/desorption)
moves have been included in the dynamics. In
the present work, we extend these numerical studies to the effects of
lateral adsorbate diffusion to nearest-neighbor positions
(corresponding to spin-exchange moves in magnetic language). 

Our work is also inspired by recent electrochemical pulse experiments on
the dynamics of the processes that occur after lifting the
$(22\times\sqrt{3})$ reconstruction of
the Au(111) surface.\cite{He01} In that study, during a short
positive pulse of the electrode potential, the reconstruction is
lifted immediately, and excess Au atoms are released onto the then ($1\times 1$) substrate,
where they quickly nucleate and form islands. After the  pulse, the
surface slowly reconstructs again, consuming the excess ions. During
this process, however, the adsorbate structure still undergoes
post-deposition dynamics, where small islands tend to decay faster,
and large islands gain in size before they eventually decay. However, we
emphasize that the
connection of our model to these experiments is only qualitative. As the
source for the adsorbate, we use ions from the solution, whose
supersaturation can be pulsed. Moreover, we
neglect the geometry of the surface and the dynamics of the lifting and
reconstruction process. Also, the peak coverages in our simulations
are much higher than in the experiment. We rather focus on the general
characteristics of the growth and subsequent decay of submonolayer
adsorbates with attractive interactions during and after a pulse of their
supersaturation, and how these characteristics compare to the KJMA theory. Though we keep
the model as simple as possible, nearest-neighbor diffusion is an
essential characteristic of the experimental system and must be
included in the dynamics of the model.

We shall not repeat here the discussion of the theory of the decay of
metastable phases (see, e.g., Refs.~\onlinecite{RikvoldBook94, Ramos99}), but only
give a brief summary. The KJMA theory in its simplest form describes
the decay of a metastable phase to an equilibrium
phase, driven by a difference in free energy, from simple basic
assumptions: negligibly small nuclei form at random and subsequently
grow isotropically to larger droplets. An extended volume is defined
as the sum of the volumes of all droplets, which grow and overlap
freely, even upon collision. Since the nucleation events are
completely random, they can also occur in the already transformed
volume. The true transformed volume is the space covered by at least
one droplet. Due to the randomness of the process,
the relation between the extended volume fraction
$\varphi_\mathrm{ext}$ and the (true)
transformed volume fraction $\varphi$ can be shown to be~\cite{Avrami40}
\begin{equation}
1-\varphi=\exp(-\varphi_\mathrm{ext}).
\label{eq:avrami}
\end{equation}
Assuming a constant radial growth velocity $v$ and a constant nucleation
rate $I$ (corresponding to continuous nucleation), the extended volume is, in two dimensions,
\begin{equation}
\varphi_\mathrm{ext}=\frac{\pi I v^2 t^3}{3},
\label{eq:extvol}
\end{equation}
at time $t$ after the beginning of the phase transformation. Continuous
nucleation is often associated with homogeneous nucleation due to
thermal fluctuations. Another case, initial nucleation, where all the
nuclei are present at the very beginning of the transformation, is
often associated with heterogeneous nucleation at defects. 
In the case of continuous nucleation, thermal fluctuations that lead to the formation of small droplets
obey metastable thermodynamics, and knowledge of the size dependence
of the free energy~$F$ of compact small droplets is needed. In a
mesoscopic description, which should hold sufficiently below the
critical temperature, this is obtained as a sum of a surface and a
volume term. In two dimensions, this becomes
\begin{equation}
F(r)=2\pi r\sigma_0 - (\mu - \mu_0)\Delta \theta \pi r^2, 
\label{eq:droplet}
\end{equation}
for a cluster with effective radius $r$ and surface
tension~$\sigma_0$ along a primitive lattice vector at equilibrium
(whose zero-field Ising analog  is known exactly~\cite{Onsager44}).
The difference in coverage $\Delta\theta$ between the two
degenerate equilibrium phases corresponds to the difference of
the spontaneous magnetization of the two degenerate phases at zero field in
the Ising model.\cite{Yang52} The use of the zero-field values is
believed to be justified, and a recent publication~\cite{Wonczak00}
supports the point of view that the surface tension is not strongly
field dependent (though that study addresses a lower
temperature). Conversely, the use of the macroscopic surface tension
for microscopic structures is questionable, and more accurate expressions
are a matter of research~\cite{Shneidman99, Shneidman01, Berthier04,
  Shneidman04} and are sought in this paper as well. The
free energy has a maximum at the critical droplet
size (volume) $s^*$. The existence of this maximum is the reason why the
phase transformation is activated. On average, when smaller than
$s^*$, the droplets tend to dissolve, and when greater than $s^*$, they tend
to grow. The critical droplet size is a statistical quantity
and is less sharply defined the flatter the shape of the maximum of the free energy. 

The velocity and the ratio of nucleation and growth are influenced by
lateral mass transport. In ultra-high vacuum (UHV) and under the
impact of a constant flux of the adsorbate, desorption may often be
neglected (irreversible growth). In electrochemistry (EC), where the control parameter is the
electrochemical potential, quasi-equilibrium is established for
the subcritical fluctuations, so that desorption has to be included in
a model description (reversible growth). As a consequence, in UHV, surface
diffusion is generally the only means of lateral mass transport, while
in EC, it can as well occur through the solution by
independent desorption and adsorption events at different locations. In surface
science, it has long been established that diffusion introduces
correlations between clusters, since it reduces the nucleation probability in the
vicinity of existing islands.\cite{Bartelt96,Ratsch03} Recent publications have addressed this
violation of the basic assumptions of KJMA kinetics.\cite{Pineda02,
  Fanfoni03} Since the number of nucleation events is effectively \emph{diminished},
phase transformation is slowed down. However, it has not been
considered that the enhanced mass transport can \emph{increase} the
nucleation and growth rates, resulting in an acceleration of phase
transformation. Thus, truly microscopic model calculations are
desirable, and we provide them in this paper. This microscopic
approach distinguishes our work from most other studies that aim at
generalizing the KJMA theory.

The rest of this paper is organized as follows: We introduce the model
and describe our model calculations
in Sec.~\ref{par:model}. Results  are presented in
Sec.~\ref{par:results}.  First, we
discuss the phase-transformation kinetics~(Sec.~\ref{par:results:kinetics})
and how they are affected by diffusion. Second, we discuss
the effects of diffusion on the size distributions of 
supercritical~(Sec.~\ref{par:results:clusterdist}) and 
subcritical~(Sec.~\ref{par:results:subcrit}) clusters. For the latter,
we give a theoretical explanation in terms of the free energy of
lattice animals.
 A summary and conclusions follow in
Sec.~\ref{par:summary}.

\section{Model and algorithm}
\label{par:model}

\subsection{Model}
\label{par:model:model}
We describe the electrochemical deposition of ions onto a metal
surface in terms of a square lattice-gas model with attractive
nearest-neighbor interactions. The grand-canonical Hamiltonian of the
system reads
\begin{equation}
\mathcal{H}=-\phi \sum_{\langle i,j\rangle}c_i(t)c_j(t)-\mu \sum_i
c_i(t),
\label{eq:hamiltonian}
\end{equation}
where $\phi$ is the interaction constant and $\mu$ the adsorbing ions'
electrochemical potential in the solution, which represents an
infinite reservoir. The $c_i$ are occupation variables at site~$i$ with the
values
\begin{equation*}
c_i(t)=\begin{cases}
0&  \text{if empty}\\
1&  \text{if occupied}.
\end{cases}
\end{equation*}
The first sum in Eq.~(\ref{eq:hamiltonian}) runs over all pairs of
nearest-neighbor sites, and the second runs over all lattice
sites. Water and counter ions are not explicitly
included. The Hamiltonian is equivalent to the Ising model, up to an
additive factor.\cite{Footnote1,Lee52a, RikvoldBook94}

We perform model simulations for an $L\times L=256\times 256$ square lattice,
applying periodic boundary conditions,
for the values of the model parameters $\phi=4$, and
$|\mu-\mu_0|=0.4$,\cite{Footnote2} where $\mu_0$ is the
electrochemical potential at coexistence, and
$\mu > \mu_0$ favors adsorption. The
temperature $T$ equals $0.8\, T_{\mathrm{c}}\approx 0.453873\,\phi\approx
1.815348\,J$,\cite{Footnote1} with $T_{\mathrm{c}}$ the critical
temperature of the Ising model for the given 
parameters.\cite{Onsager44} Energy and temperature units are chosen such that
Boltzmann's constant $k_{\mathrm{B}}=1$.

\subsection{Algorithm}
\label{par:model:calc}
To investigate the kinetics of the phase transformation, we perform
kinetic Monte Carlo simulations,\cite{BinderBook97} where the
transition probability for an elementary step is given by the Glauber
acceptance probability~\cite{Glauber63}
\begin{equation}
P=\frac{1}{1+\exp \left[(E_f-E_i)/k_{\mathrm{B}} T\right]},
\end{equation}
with $E_f$ and $E_i$ the configurational energies in the final
and the initial configurations, respectively. We choose this dynamic
for the sake of consistency with earlier work.\cite{Ramos99} Other
dynamics that are commonly used in studies of electrochemical
adsorption, such as the one-step dynamic~\cite{Kang89, Fichthorn91}
and the transition dynamic algorithm~\cite{Ala-Nissila92,
  Ala-Nissila92a} also contain a \emph{local} transition barrier
and are commonly known as Arrhenius dynamics. For further discussion
and applications of such dynamics, see, e.g., Refs.~\onlinecite{BrownBook99,
  Mitchell01, Hamad04, Buendia04}.

Two kinds of elementary steps are considered: an adsorption/desorption
step, which corresponds to a non-conserved order parameter, and a
diffusion step to a nearest-neighbor position, which conserves the
order parameter.\cite{Gunton83} A new configuration is chosen by
randomly picking a lattice site and proposing a trial step: either, with probability $1/(R+1)$, an
adsorption/desorption step or, with probability $R/(R+1)$, a
diffusion step to a randomly chosen nearest-neighbor
position. Thus, the ratio of the diffusion attempt probability to the
adsorption/desorption attempt probability equals $R$. Explicitly,
we propose adsorption when we choose an empty site for an
adsorption/desorption step, and desorption when we choose an occupied
site -- there is no second-layer growth. When we choose an empty
site for a diffusion step, the
initial configuration is maintained -- vacancies do not diffuse. Only
after attempting an adsorption/desorption step is the simulation clock
updated by one Monte Carlo step (mcs); hence, a higher diffusion rate $R$ does not slow down the
time scale for adsorption/desorption, which is the physically relevant
time scale for the phase transformation. After $L^2$ mcs, the
simulation time thus has increased by one \emph{Monte Carlo step per
  site}~(MCSS). Hence, one MCSS means in our simulations one
adsorption/desorption attempt per site.

We start our simulations with an empty lattice, set the electrochemical
potential to $\mu=\mu_0+0.4$, and let the system relax toward the stable
configuration. We sample the time development of the coverage~$\theta$ (the
appropriate order parameter of the system, equivalent to $\varphi$ in
KJMA theory) and, at certain coverages,
evaluate the cluster-size distribution, using the Hoshen-Kopelman
algorithm.\cite{Hoshen76} We perform model calculations for various
values of $R$ from 0 to 1000. Every calculation is the average over
100 independent simulation runs.

One of our goals is the understanding of the electrochemical deposition and subsequent
dissolution of metal adsorbates after short pulses of the
electrode potential. To that end, in addition to the
constant-potential simulations described in the previous paragraph, we
perform potential-pulse simulations for values of $R$ from $0$ to
$10\,000$ starting
from a metastable low-coverage configuration (result of an equilibration at
$\mu=\mu_0-0.4$), set $\mu=\mu_0+0.4$ and wait until a certain
coverage $\theta_s$
is reached. Then, we switch back to $\mu=\mu_0-0.4$, at which the adsorbate
desorbs again, and let the system equilibrate.

\section{Results}
\label{par:results}
\begin{figure}
\includegraphics[clip]{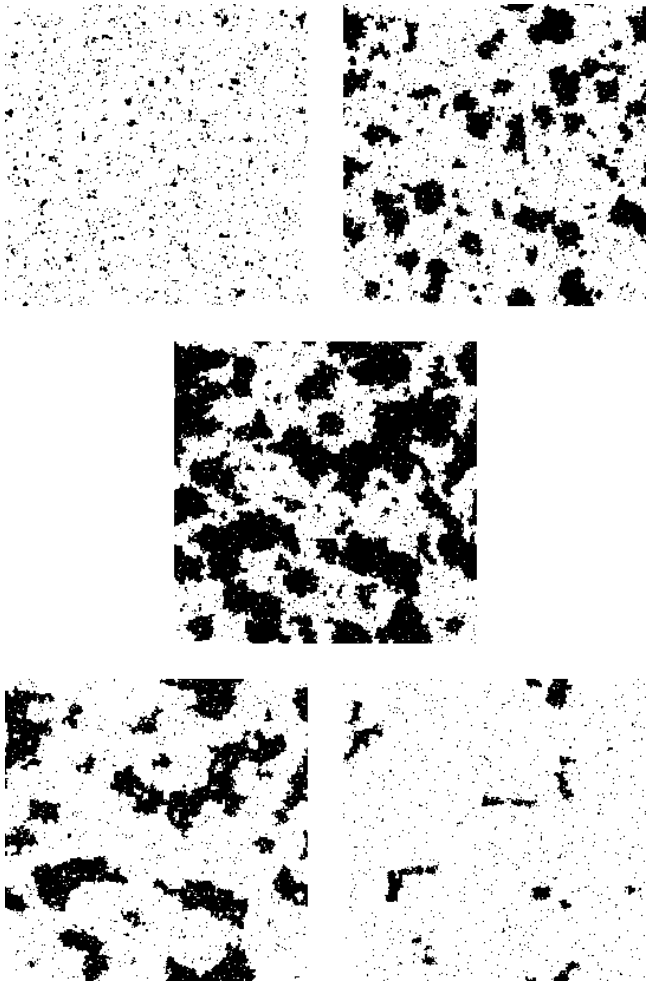}
\caption{\label{fig:snapshot_pulse}Snapshots at different stages of the phase transformation
  during the potential-pulse simulations. $R=0$. Coverages (from upper
  left to lower right): $\theta=0.05$, $\theta=0.25$ and $\theta=0.5$
  during the adsorption step, $\theta=0.25$ and $\theta=0.05$ during
  the desorption step.}
\end{figure}

In this section, we present the results of our model simulations for
the constant-potential and the potential-pulse cases. 
Snapshots of the adsorbate morphology at different time
steps~(Fig.~\ref{fig:snapshot_pulse}) show the nucleation, growth, and coalescence
of clusters of quite irregular shapes (a more detailed discussion of
this figure follows in Sec.~\ref{par:results:clusterdist:nodiff}). It has been established, however,
that the classical KJMA theory gives a good statistical description of
phase transformation with microscopic dynamics, at least at its earlier stages and in the
absence of diffusion, even though it assumes the growth of circular
droplets.\cite{Ramos99} Despite that the KJMA theory does not make
assumptions about the nucleation and growth mechanisms, it is not
clear how far it will hold when adsorbate diffusion is
included in the kinetics of the lattice-gas model. We shall address
this question in the following.

\subsection{Phase-transformation kinetics}
\label{par:results:kinetics}
\begin{figure}
\includegraphics[clip]{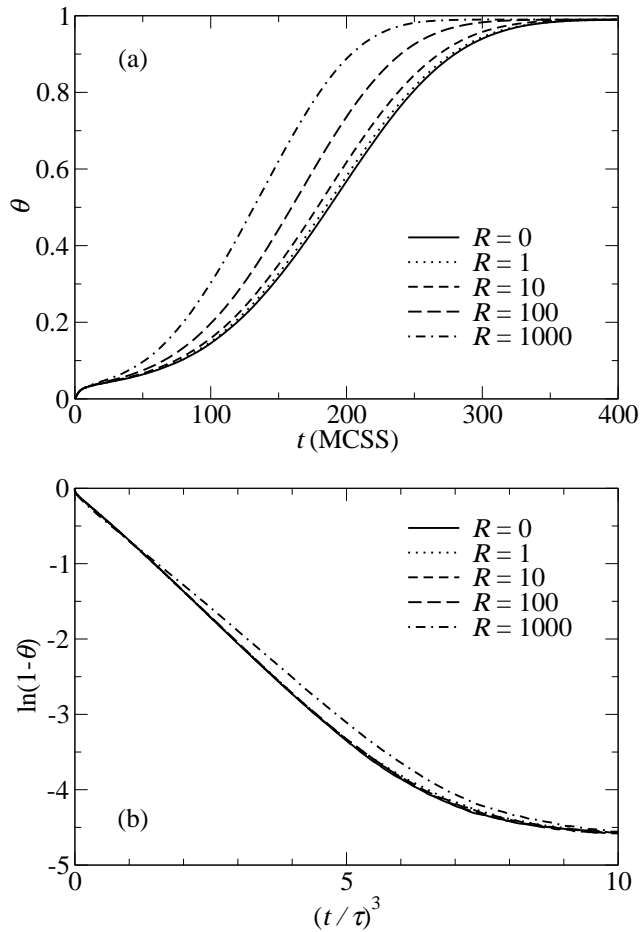}
\caption{\label{fig:coverage_avrami_nopulse}Time development of the coverage during constant-potential
  simulations. 
  (a) Linear plot vs time. (b) Logarithmic plot of
  fraction of uncovered area vs cube of normalized time.}
\end{figure}

Monte Carlo results for the time development of the coverage at
constant potential~[Fig.~\ref{fig:coverage_avrami_nopulse}(a)] apparently show
the shape familiar from KJMA theory,\cite{Johnson39} but with an
acceleration of the phase transformation due to diffusion.
For a diffusion rate of $R=10$ the effect is not very marked, but
it becomes quite pronounced for higher $R$. In order to quantify the
effect, we define the metastable lifetime $\tau$ as the time when the
coverage reaches $\theta=1/2$. The results are shown in
Table~\ref{tab:lifetime}. The lifetime $\tau$ decreases weakly from
$R=0$ to $R=10$, and then more strongly for larger $R$.

\begin{table}
\caption{\label{tab:lifetime}Metastable lifetime for different diffusion rates~$R$:
 $\tau_{\theta}$ from $\theta=0.5$ in constant-potential simulations  and
 $\tau_p$ from peak position in potential-pulse simulations.} 
\begin{ruledtabular}
\begin{tabular}{lll}
$R$ & $\tau_{\theta}$ (MCSS) & $\tau_p$ (MCSS) \\ \hline
0     & 187 & 184 \\
1     & 185 & 184 \\
10    & 179 & 177 \\
100   & 160 & 160 \\
1000  & 132 & 130 \\
10\,000 &     & 111 \\
\end{tabular}
\end{ruledtabular}
\end{table}

KJMA theory predicts a linear dependence of the logarithm of the area
fraction of the untransformed phase ln$(1-\theta)$ on $t^n$ with the
Avrami exponent $n=3$~[Eqs.~(\ref{eq:avrami}) and (\ref{eq:extvol})]. In
order to distinguish a change in the linear dependence from the
acceleration of the phase transformation, we normalize the time scales to the metastable
lifetime~$\tau$. Figure~\ref{fig:coverage_avrami_nopulse}(b) 
shows almost
perfect data collapse up to $R=100$. At larger $R$, the slope is
initially steeper. This corresponds to a relatively higher increase of
the transformation rate, which levels off at $t\approx\tau$. To obtain
quantitative information on the kinetics, we perform for all values of
$R$ unweighted least-squares
fits to straight lines of the corresponding data points 
with the time scale unnormalized. The slopes give the product
$Iv^2$. From Eqs.~(\ref{eq:avrami}) and (\ref{eq:extvol}), the
contributions of the nucleation and growth rates, $I$ and $v$, to the
acceleration of the phase transformation cannot be distinguished.  For a
separate estimate of $I$ and $v$, one needs to know the variance of
the order parameter,\cite{Ramos99} which we did not sample in our
simulations.  

The fits
have to be restricted to a certain time interval. Around $t=\tau$, the
plots show a slight kink, a result of surface tension effects around the
percolation coverage, which disturb the independent growth of
the droplets. Therefore, we choose $0.9\times \tau$ as the upper limit. At
very short times the system relaxes quickly to the metastable 
state. Therefore, we choose the lower time limit by
stepwise decreasing it until $\chi^2$ per degree of freedom starts to
increase steeply. We here refrain from correcting for the background of the
metastable phase, in contrast to Ref.~\onlinecite{Ramos99}. 

The results for $Iv^2$ shown in
Table~\ref{tab:avramifits} reflect the acceleration of the phase
transformation. The product $Iv^2$ is inversely proportional to the
cube of the metastable lifetime~$\tau$, and as long as the dynamics of
the phase transformation do not change qualitatively, the situation is
tantamount to a mere rescaling of the time scale. Then, the film
morphology at the same coverage should not depend on $R$. 
The product $Iv^2 \tau^3$ is constant for $R\le100$, but differs at
$R=1000$, indicating a qualitative change in the phase-transformation
dynamics for large $R$. Accordingly, we find that the curve for the same $R$ fits
less well to a straight line, so that the fitting interval is much
shorter. We will return to this point in the next paragraph. In
Sec.~\ref{par:results:clusterdist:diff} we will show that this change coincides
with a change of the film morphology.

\begin{table}
\caption{\label{tab:avramifits}Kinetic parameters of the KJMA theory from fits of the phase
 transformation plots similar to Fig.~\ref{fig:coverage_avrami_nopulse}(b) for
 different diffusion rates $R$. $I$
 nucleation rate, $v$ linear growth rate, $\tau$ metastable life
 time.}
\begin{ruledtabular}
\begin{tabular}{lll}
$R$   & $Iv^2$ ($\mathrm{MCSS}^{3}$) & $Iv^2\tau^{3}$\\ \hline
0     & $9.1\times10^{-8}$ & 0.59\\
1     & $9.5\times10^{-8}$ & 0.59\\
10    & $1.0\times10^{-7}$ & 0.59\\
100   & $1.5\times10^{-7}$ & 0.59\\
1000  & $2.5\times10^{-7}$ & 0.56\\
\end{tabular}
\end{ruledtabular}
\end{table}

\begin{figure}
\includegraphics[clip]{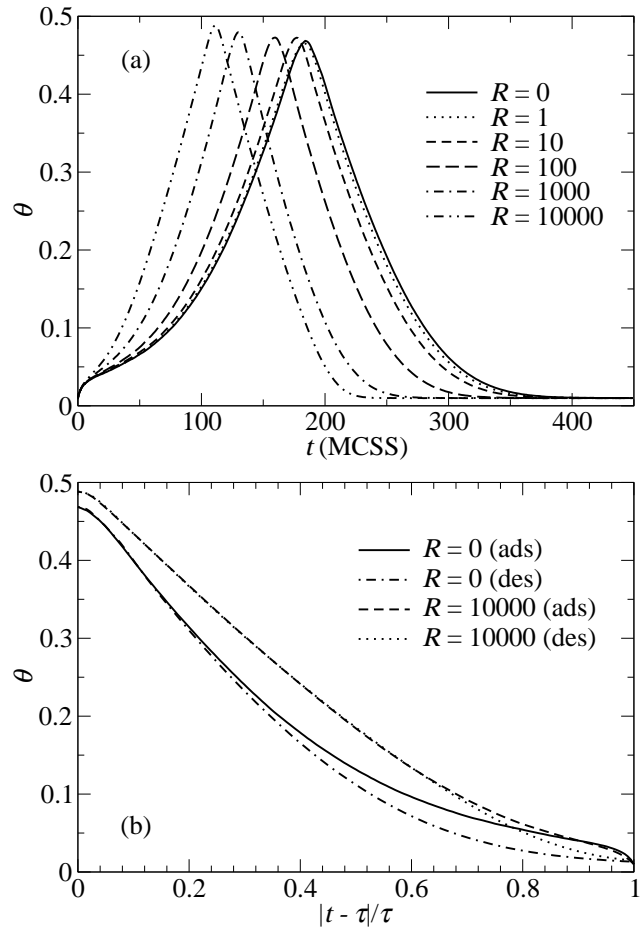}
\caption{\label{fig:coverage_pulse}Time development of the coverage during
  potential-pulse simulations with switching coverage $ \theta_s=0.5$. (a) Linear. (b) Mirrored
  at $\tau$ and normalized with~$\tau$. Curves labeled (ads) during adsorption step, curves
  labeled (des) during desorption step.}
\end{figure}

\begin{figure}
\includegraphics[clip]{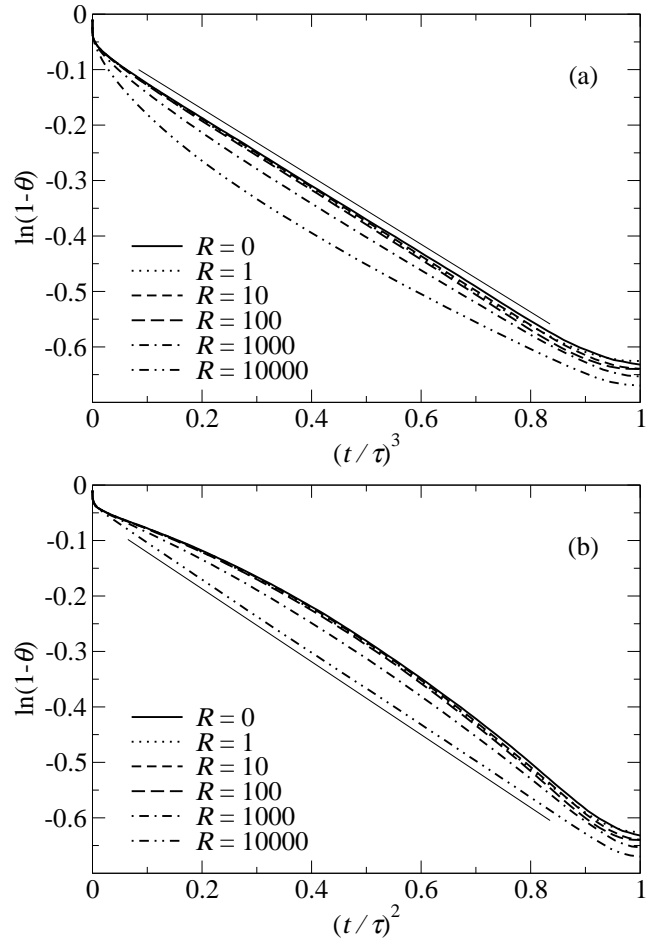}
\caption{\label{fig:avrami_pulse}Time development of the coverage during potential-pulse
  simulations up to $\tau$. Logarithmic plot of fraction of uncovered
  area vs (a) cube (b) square
  of normalized time. The thin straight lines are guides to the eye.}
\end{figure}

The adsorption step of the potential-pulse calculations is equivalent
to the constant-potential simulations. After the potential switch, the
adsorbate desorbs, resulting in peak-shaped coverage curves, as we demonstrate
for a switching coverage of $0.5$ in
Fig.~\ref{fig:coverage_pulse}(a). The peak position gives another estimate for the metastable
lifetime~$\tau$ (Table~\ref{tab:lifetime}). It coincides with the
former estimate. In the absence of diffusion, the desorption is faster
than the adsorption -- the desorption can proceed without previous
nucleation by simple shrinking of the clusters. The asymmetry in the
kinetics between adsorption and desorption decreases with increasing
$R$ and is very small at
$R=10\,000$~[Fig.~\ref{fig:coverage_pulse}(b)].
At this diffusion rate, the decay of the metastable phase seems to be
much better described by an Avrami exponent $n=2$, consistent with
instantaneous nucleation (Fig.~\ref{fig:avrami_pulse}).

During the growth of the adsorbate
phase, nucleation may be effectively inhibited in the \emph{capture
  zones} around clusters, and the size of these zones depends on the
diffusion rate.\cite{Evans02, Family02} Concerning the apparent Avrami exponent $n=2$
for $R=10\,000$, one may speculate that nucleation is initially very
fast, but gets inhibited quickly on the whole surface, which is
completely covered by the capture zones of existing clusters. The
result is instantaneous, though homogeneous nucleation. The
subsequent adsorption kinetics are governed solely by the
growth rate, as is the case for the desorption kinetics after the
potential step. This explains the almost symmetric coverage curves for
$R=10\,000$ [Fig.~\ref{fig:coverage_pulse}(b)].

\subsection{Cluster-size distributions}
\label{par:results:clusterdist}

We characterize the film morphology by means of the cluster-size
distributions. We show the number
densities, i.e. the total number
of clusters of size $s$ on the finite lattice, divided by the number
of lattice sites. This
equals the probability of finding on a site the center of a cluster of
size $s$.  The size $s$ is the number of particles in the cluster. Due to the
low number densities for large cluster sizes, we use a
growing bin size for the histograms, resulting in an equal
distribution of data points on a logarithmic scale. We show results
for the potential-pulse simulations.

\subsubsection{In the absence of diffusion}
\label{par:results:clusterdist:nodiff}

\begin{figure}
\includegraphics[clip]{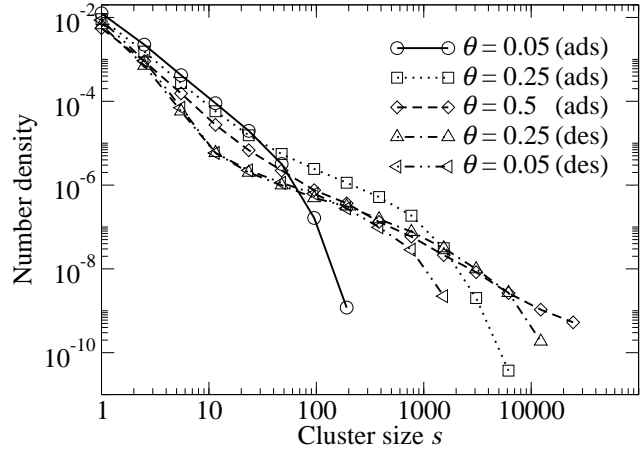}
\caption{\label{fig:histo_M0.R0}Cluster-size distributions (number of clusters of size $s$
  per unit area) for potential-pulse simulation
  during the adsorption step (ads) and the desorption step
  (des). Switching coverage $\theta_s=0.5$, diffusion rate $R=0$.}
\end{figure}

In the absence of diffusion, we show the evolution of the cluster-size
distributions at different stages of the course of the potential-pulse simulations as
log-log plots in Fig.~\ref{fig:histo_M0.R0}. During the
adsorption step, the maximum cluster size increases with time, as
existing clusters grow to larger sizes. The number densities of the 
cluster sizes that had been present previously decrease. This indicates that
with time fewer clusters nucleate and grow into a certain size interval
than grow out of it. In the early stages of the desorption step, it can be clearly
seen that, while the densities of the largest and of medium-sized and
small clusters are decreasing, some intermediately large clusters
(around $s=1000$) still
gain in population. This is also true in the presence of diffusion
(not shown). Later, the distribution gets consumed from the large-cluster
side. Hence, the distributions at equal coverage differ
markedly between the adsorption and the desorption step. This difference
is clearly reflected in the snapshots of the film morphology in
Fig.~\ref{fig:snapshot_pulse}, when comparing equal coverages during
adsorption and desorption. The latter show fewer, but larger
clusters. These snapshots also demonstrate the effect of coalescence during
adsorption on the morphology -- individual structures that form larger
aggregates at later times can be clearly identified. Conversely, during
desorption, dissociation of smaller clusters from larger aggregates
may occur. A complete theory of cluster-size distributions, which we do not aim
to present here, must consider these complications. We conclude that,
during adsorption, existing clusters grow to larger sizes and may
undergo coalescence, while new and thus smaller clusters can
only appear on the untransformed area, thus with reduced
probability. In the early stages of
desorption, ripening of the morphology occurs in parallel with a decrease of
the coverage, while later, clusters of all sizes shrink at constant
rates. 

\subsubsection{In the presence of diffusion}
\label{par:results:clusterdist:diff}

\begin{figure}
\includegraphics[clip]{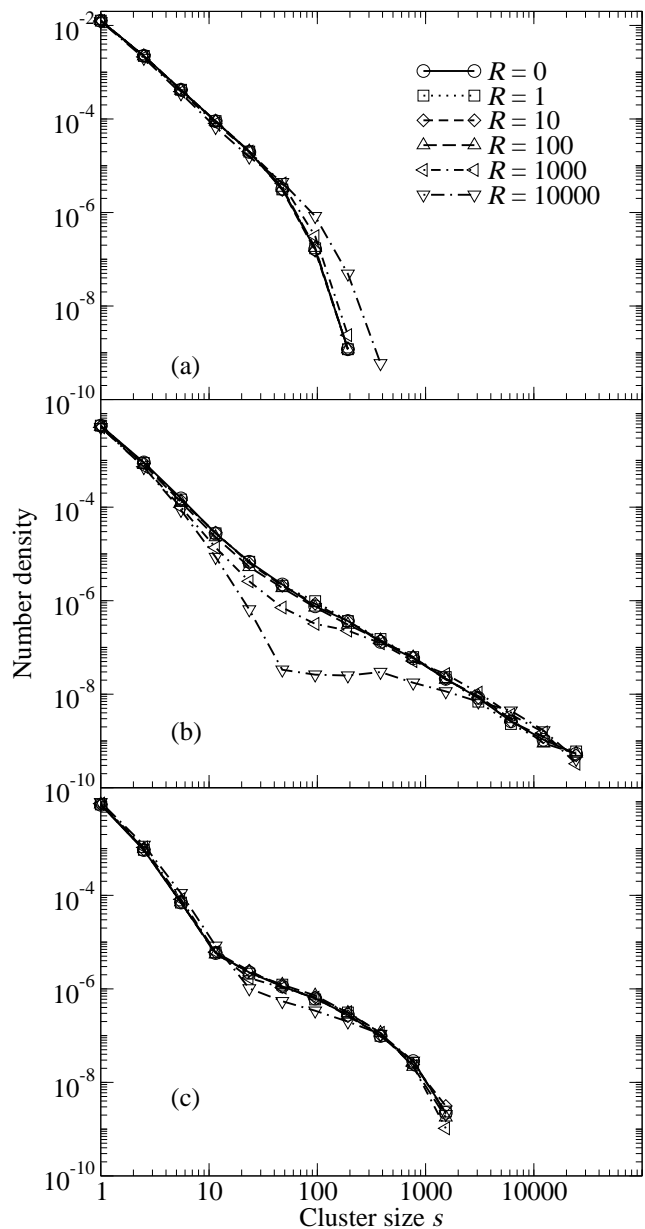}
\caption{\label{fig:histo_M0.time}Cluster-size distributions (number of clusters of size $s$
  per unit area) for potential-pulse simulations
  with switching coverage $\theta_s=0.5$. (a) $\theta=0.05$ (during adsorption step). (b)
  $\theta=0.5$ (potential switch). (c) $\theta=0.05$ (during desorption step).}
\end{figure}

\begin{figure}
\includegraphics[clip]{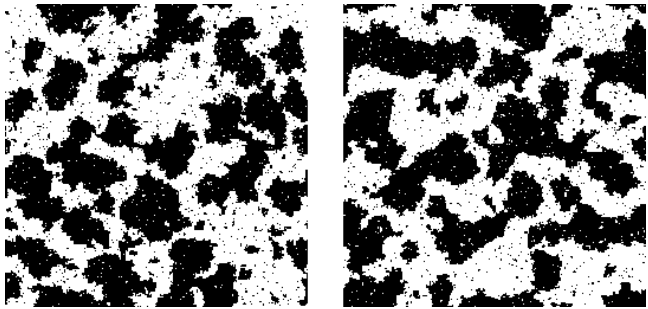}
\caption{\label{fig:snapshot_diff}Snapshots  for diffusion rates
  $R=0$ (left) and $R=1000$ (right). In both cases, $\theta=0.5$.}
\end{figure}

For the potential-pulse simulations in the presence of diffusion,
results for different diffusion rates at $\theta=0.05$ during adsorption,
$\theta=0.5$ (the switching coverage), and $\theta=0.05$ during desorption
are shown in Fig.~\ref{fig:histo_M0.time} as log-log
plots. In the early stages of growth, we find that for fast diffusion
rates ($R\geq 1000$), the maximum cluster sizes present are
larger~[Fig.~\ref{fig:histo_M0.time}(a)]. At later stages, there is a
marked reduction through fast diffusion of the densities of medium-sized clusters,
together with an increase of the densities of large
clusters~[Fig.~\ref{fig:histo_M0.time}(b)]. In accordance with this
observation, snapshots of the phase transformation show a coarser
morphology for $R=1000$ than for $R=0$ (Fig.~\ref{fig:snapshot_diff}). Near the end of
the reverse step, the reduction of the amount of medium-sized
clusters prevails, but much weaker, such that the morphologies for
different $R$ get more similar during
desorption~[Fig.~\ref{fig:histo_M0.time}(c)]. For $R \leq 100$,
the distributions are virtually indistinguishable from the case
without diffusion. This indicates that the phase-transformation
dynamics are similar and just proceed on a faster time 
scale~(Sec.~\ref{par:results:kinetics}).

We find the reduction of medium-sized clusters more likely to be a result of
increased coalescence rates than of a change of the radius-dependent
growth law of the clusters. While the former can be a result of a
collective motion of clusters established by edge diffusion or of
dissociation and diffusion of monomers and is well established in
surface science,\cite{Wen94, Pai97, Khare96} the latter would imply an unphysical
slowing-down of growth only at medium sizes.

We have seen that diffusion affects the cluster-size distributions
only at diffusion rates high enough that also the phase-transformation
kinetics cross over from continuous to initial nucleation, $R\ge1000$ 
(Sec.~\ref{par:results:kinetics}). Lower diffusion rates speed up the
dynamics without changing them qualitatively. A rough estimation
considering the number densities of monomers suggests that at these
lower rates, diffusion is still not more frequent than
adsorption/desorption, as the mere numerical values of $R$ na\"{\i}vely might suggest.

\subsection{Subcritical cluster distributions}
\label{par:results:subcrit}
Unlike the supercritical droplets, whose size distribution is governed
by the phase-transformation dynamics, the 
subcritical droplets can be considered to be quasi-equilibrium
fluctuations. This allows a thermodynamic approach for a theoretical
description of their size distribution. We first show simulation
results of the size distributions. Second, we calculate their
theoretical quasi-equilibrium distribution within a lattice-animal
model.

\subsubsection{Simulation results}
\label{par:results:subcrit:sim}
\begin{figure}
\includegraphics[clip]{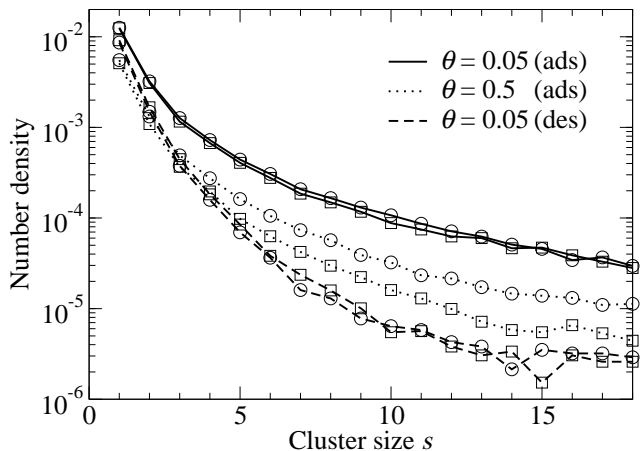}
\caption{\label{fig:subcrit_M0}Subcritical cluster-size distributions (number of clusters of size $s$
  per unit area) for $R=0$ (circles) and
  $R=1000$ (squares) at different stages of the potential-pulse
  simulations. Curves labeled (ads) during the adsorption step, curves
  labeled (des) during the desorption step.}
\end{figure}

We show the number densities, as explained in
Sec.~\ref{par:results:clusterdist}, of the subcritical clusters for
potential-pulse simulations with a switching coverage of
$\theta_s=0.5$ in Fig.~\ref{fig:subcrit_M0}. At all stages of the simulation, the size
distribution of small clusters decreases monotonically with increasing
size. The distribution develops with time. During the
adsorption step, the number densities decrease. This decrease is
somewhat greater than expected from the decrease of the untransformed
surface fraction. Column~4 in Table~\ref{tab:monomerreduction} shows
the fraction to which the free surface gets reduced compared to
$\theta=0.05$, and column~5 shows the corresponding reduction of the
monomer density. The reduction of the monomer density is always
greater than the reduction of the free surface, and it is
enhanced by increasing diffusion rate. This indicates that in a capture zone close to an existing cluster
nucleation is suppressed, since aggregation to the cluster is more
likely. The size of the capture zones is increased by diffusion. During the desorption
step, the number densities of small clusters are generally further
diminished (with the exception of monomers and dimers, whose densities
increase). Again, in the course of desorption, the distributions with
and without diffusion approach each other. 

\begin{table}
\caption{\label{tab:monomerreduction}Reduction of the monomer number density $n_1$ relative to the
 untransformed surface fraction. $R$ diffusion rate, $\theta$
 coverage. See Sec.~\ref{par:results:subcrit:sim}.}
\begin{ruledtabular}
\begin{tabular}{lllll}
$R$ & $\theta$ & $n_1(\theta)$ & $ \frac{1-\theta}{1-0.05}$ &
   $\frac{n_1(\theta)}{n_1(0.05)} $\\ \hline
$0$ & $0.05$ & $1.25\times10^{-2} $ & $ 1$ & $1 $ \\
& $0.25$ & $9.10\times10^{-3} $ & $0.79 $ & $0.72 $ \\
& $0.5$ & $5.53\times10^{-3} $ & $0.53 $ & $ 0.44$ \\
$1000$ & $0.05$ & $1.24\times10^{-2} $ & $1 $ & $1 $ \\
& $0.25$ & $8.60\times10^{-3} $ & $ 0.79$ & $ 0.69$ \\
& $0.5$ & $5.51\times10^{-3} $ & $0.53 $ & $0.41 $ \\
\end{tabular}
\end{ruledtabular}
\end{table}

\subsubsection{Lattice-animal model}
\label{par:results:subcrit:animals}
\begin{figure}
\includegraphics[clip]{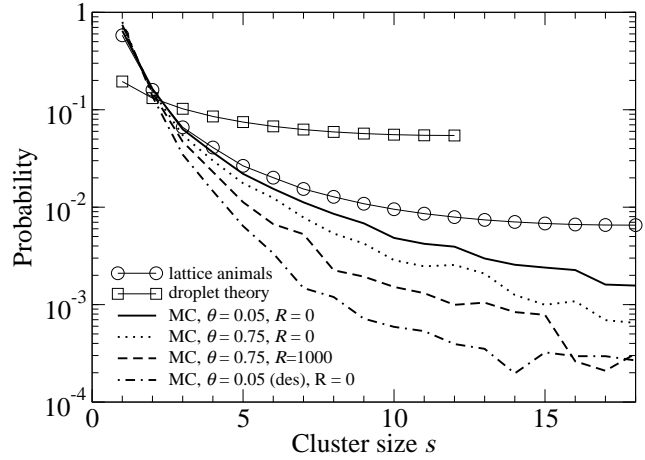}
\caption{\label{fig:subcritical_theory}Subcritical cluster-size probability distributions
  (probability that a subcritical cluster is of size $s$):
  Monte-Carlo results and theoretical predictions from droplet theory
  and from counting lattice animals. Curve labeled (des) during the
  desorption step, others during the adsorption step.}
\end{figure}

\begin{figure}
\includegraphics[clip]{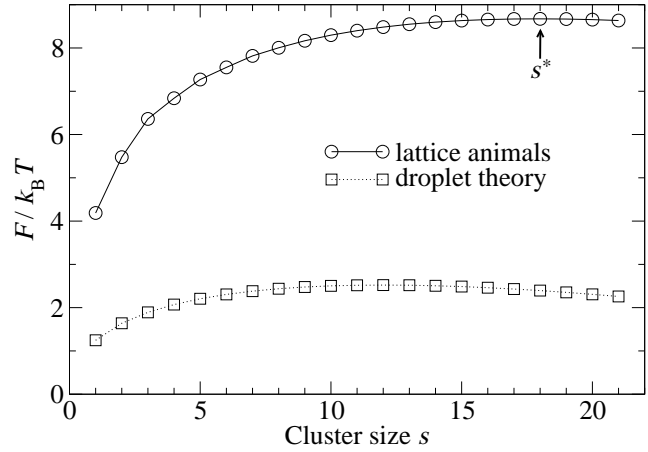}
\caption{\label{fig:freeener}Free energy of small clusters according to droplet theory and
  from counting lattice animals. The critical cluster size is
  labeled $s^*$.}
\end{figure}

The cluster-size probability distribution gives the probability $P(s)$ that a
subcritical cluster is of size $s$, and it is normalized such that the
sum of all probabilities equals~1. To calculate it, we need an expression for the
free energy $F(s)$ of a cluster of size $s$, since
\begin{equation}
P(s) \propto \exp (-F(s)/k_{\mathrm{B}}T).
\end{equation}
The theoretical prediction from droplet theory [Eq.~(\ref{eq:droplet})], in two dimensions,
using  $\sqrt{s/\pi}$ as an effective
radius $r$ and 
the exact zero-field Ising values for $\sigma_0$ and
$\Delta\theta$, is shown in Fig.~\ref{fig:subcritical_theory}. The agreement is
rather poor. Since the surface tension, which is a macroscopic quantity,
may not be a good description for such small clusters, we seek a
better statistical-mechanical expression for the cluster free
energy. If we knew all possible configurations of a cluster of size
$s$ and their numbers of bonds to nearest-neighbor unoccupied
sites (including holes) $t$,
we could compute its restricted partition function exactly as 
\begin{equation}
Z(s)=\exp\left(\frac{s\,\left(\mu-\mu_0\right)}{k_{\mathrm{B}}T}\right) 
\sum_t w(s,t)\exp \left(-\frac{t\phi}{2\,k_{\mathrm{B}}T}\right).
\end{equation}
More specifically, we need to know the number $w(s,t)$ of degenerate clusters with same $s$ and
$t$. These numbers are very high unless $s$ is very small. As a
consequence, even with the help of computers, they are accessible only
up to moderate values of $s$. Generating these configurations is known
as counting lattice animals.\cite{GolombBook65, Stauffer79,
  Redelmeier81, Jensen01, Shneidman04} 
In the lattice-animal literature (see, e.g., Ref.~\onlinecite{Sykes76}), the
perimeter, if considered, is usually defined as the number of unoccupied
nearest-neighbor sites and not as the number of bonds to these sites,
and only $w(s)$ is known. For the calculation of $w(s,t)$, we use the
algorithm of Redner.\cite{Redner82} We show the results for animals
up to $s=21$ in Table~\ref{tab:animals}. The computation took about 65
hours of CPU time on an Intel Xeon-based workstation. The $w(s)$ from our calculations
are in accord with the known results.\cite{Redelmeier81, Jensen01} 
From the partition function, the free energy is calculated as
\begin{equation}
F(s)=-k_{\mathrm{B}}T\ln Z(s),
\end{equation}
so that
\begin{equation}
P(s) = Z(s) /\sum_{s=1}^{s^*} Z(s),
\end{equation}
summing up to the critical cluster size $s^*$. This calculation, as the droplet theory, assumes
non-interacting subcritical clusters which are in quasi-equilibrium
throughout the course of the phase transformation, an assumption that
should be better fulfilled at higher temperatures, as is the case in
our simulations.
The resulting cluster-size probability distribution is shown and compared to
simulation results in Fig.~\ref{fig:subcritical_theory}. At early
times and in the absence of diffusion, the agreement is
reasonable (note the logarithmic scale). The trends that we observed
for the number densities are reflected as an increasing reduction of
the probabilities with increasing cluster size for later times, in the
presence of diffusion, and during desorption, respectively. We can attribute
these trends to increasing deviations from the quasi-equilibrium
distribution -- the larger subcritical clusters are consumed faster
than they are produced, or during desorption, the small clusters are
produced faster than they disappear (remember their increase in
absolute density). In contrast, the predictions of droplet
theory are way off. This is a consequence of the surface-tension
contribution to the free energy of the clusters
(Fig.~\ref{fig:freeener}); surface fluctuations from the continuum
description, which contribute to the partition function, are
unphysical on the discrete lattice for small clusters. In the
lattice-animal model, the maximum of the free energy is shifted to larger
clusters compared to droplet theory, thus giving a critical cluster
size of $18$ instead of $12$. In both cases, the free energy changes
very little in the neighborhood of the critical cluster. An approach
analogous to the one presented in this section was independently
developed in Ref.~\onlinecite{Shneidman04} (with $w(s,t)$ computed up
to $s=17$).

\begingroup
\squeezetable
\begin{table*}
\caption{\label{tab:animals}Number $w$ of configurations of lattice animals of size $s$
 and perimeter $t$. For the definition of the perimeter, see
 Sec.~\ref{par:results:subcrit:animals}.}
\scriptsize
\begin{ruledtabular}
\begin{tabular}{lll|lll|lll|lll}
$s$ & $t$ & $w(s,t)$ &$s$ & $t$ & $w(s,t)$ &$s$ & $t$ & $w(s,t)$ &$s$ & $t$ & $w(s,t)$\\
\hline
1 &  4 &       1 & 12 & 16 &          151 & 16 & 24 &      303\,068 & 19 & 32 &     161\,217\,996\\
2 &  6 &       2 &    & 18 &         2086 &    & 26 &   1\,563\,218 &    & 34 &     506\,666\,828\\
3 &  8 &       6 &    & 20 &      13\,034 &    & 28 &   6\,095\,764 &    & 36 &  1\,230\,292\,398\\
4 &  8 &       1 &    & 22 &      58\,742 &    & 30 &  18\,173\,796 &    & 38 &  2\,044\,899\,048\\
  & 10 &      18 &    & 24 &     163\,494 &    & 32 &  36\,285\,432 &    & 40 &  1\,946\,892\,842\\
5 & 10 &       8 &    & 26 &     268\,352 &    & 34 &  42\,120\,340 & 20 & 18 &                 2\\
  & 12 &      55 & 13 & 16 &           68 & 17 & 18 &            88 &    & 20 &               426\\
6 & 10 &       2 &    & 18 &         1392 &    & 20 &          3010 &    & 22 &           12\,456\\
  & 12 &      40 &    & 20 &      11\,789 &    & 22 &       36\,112 &    & 24 &          152\,400\\
  & 14 &     174 &    & 22 &      63\,256 &    & 24 &      276\,464 &    & 26 &       1\,262\,276\\
7 & 12 &      22 &    & 24 &     250\,986 &    & 26 &   1\,603\,984 &    & 28 &       8\,194\,358\\
  & 14 &     168 &    & 26 &     633\,748 &    & 28 &   7\,477\,928 &    & 30 &      44\,227\,470\\
  & 16 &     570 &    & 28 &     942\,651 &    & 30 &  26\,922\,156 &    & 32 &     197\,485\,313\\
8 & 12 &       6 & 14 & 16 &           22 &    & 32 &  74\,496\,544 &    & 34 &     731\,782\,776\\
  & 14 &     134 &    & 18 &          864 &    & 34 & 139\,297\,108 &    & 36 &  2\,166\,643\,248\\
  & 16 &     677 &    & 20 &         9354 &    & 36 & 150\,682\,450 &    & 38 &  4\,965\,061\,010\\
  & 18 &    1908 &    & 22 &      62\,396 & 18 & 18 &            30 &    & 40 &  7\,822\,910\,077\\
9 & 12 &       1 &    & 24 &     297\,262 &    & 20 &          1728 &    & 42 &  7\,027\,047\,848\\
  & 14 &      72 &    & 26 &  1\,056\,608 &    & 22 &       26\,906 & 21 & 20 &               187\\
  & 16 &     656 &    & 28 &  2\,448\,760 &    & 24 &      236\,256 &    & 22 &              7648\\
  & 18 &    2708 &    & 30 &  3\,329\,608 &    & 26 &   1\,559\,888 &    & 24 &          115\,504\\
  & 20 &    6473 & 15 & 16 &            6 &    & 28 &   8\,193\,956 &    & 26 &       1\,068\,896\\
10 & 14 &      30 &    & 18 &          456 &    & 30 &  35\,016\,382 &    & 28 &       7\,664\,819\\
  & 16 &     482 &    & 20 &         7036 &    & 32 & 117\,417\,380 &    & 30 &      45\,205\,960\\
  & 18 &    3008 &    & 22 &      54\,908 &    & 34 & 303\,516\,966 &    & 32 &     225\,345\,274\\
  & 20 & 10\,724 &    & 24 &     317\,722 &    & 36 & 534\,018\,776 &    & 34 &     941\,355\,540\\
  & 22 & 22\,202 &    & 26 &  1\,359\,512 &    & 38 & 540\,832\,274 &    & 36 &  3\,281\,439\,844\\
11 & 14 &       8 &    & 28 &  4\,401\,192 & 19 & 18 &             8 &    & 38 &  9\,193\,698\,004\\
  & 16 &     310 &    & 30 &  9\,436\,252 &    & 20 &           914 &    & 40 & 19\,960\,812\,568\\
  & 18 &    2596 &    & 32 & 11\,817\,582 &    & 22 &       18\,756 &    & 42 & 29\,902\,719\,200\\
  & 20 & 13\,456 & 16 & 16 &            1 &    & 24 &      194\,614 &    & 44 & 25\,424\,079\,339\\
  & 22 & 42\,012 &    & 18 &          218 &    & 26 &   1\,427\,768 &    &    &                  \\
  & 24 & 76\,886 &    & 20 &         4748 &    & 28 &   8\,446\,952 &    &    &                  \\
12 & 14 &       2 &    & 22 &      46\,352 &    & 30 &  40\,680\,552 &    &    &                  \\
\end{tabular}
\end{ruledtabular}
\end{table*}
\endgroup

\section{Summary and Conclusions}
\label{par:summary}
We have performed kinetic Monte Carlo simulations of a square
lattice-gas model using Glauber dynamics to investigate the decay of a
metastable low coverage phase. The model, which contains adsorption/desorption and
nearest-neighbor diffusion moves, is used to qualitatively describe
the electrosorption and subsequent dissolution of metal ions onto a
metal substrate in potential-pulse experiments. We have observed the
phase-transformation kinetics and the size distributions of sub- and
supercritical clusters. The results have been compared to the KJMA theory. 

Nearest-neighbor diffusion speeds up the phase transformation. At low
and moderate diffusion rates, the kinetics are well described by the KJMA
theory for continuous nucleation. At high diffusion rates, we
find a crossover to kinetics that resemble initial nucleation. In the
former case, the cluster-size distributions are virtually
indistinguishable from each other, but in the latter case, the densities of
medium-sized clusters get reduced in favor of large clusters. At the same
coverage, the morphology is coarser during the desorption step than
during the adsorption step. The crossover to initial nucleation is
likely to be a consequence of the diffusion-rate dependence of the
size of the capture zones of clusters. When the diffusion rate increases, the
capture zones get larger, and for very fast diffusion rates, they
might become space-filling already in early stages of the phase
transformation.

We have calculated the free energies of subcritical clusters from their exact
restricted partition functions, which we obtain from counting lattice
animals. This enables us to calculate a theoretical prediction of the subcritical cluster-size
probability distribution. The prediction is in reasonable agreement
with simulation results, apart from some well understandable
deviations from the quasi-equilibrium distributions. In contrast,
classical droplet theory fails, since the macroscopic surface tension
is a poor description for the discrete microscopic clusters. 

Berthier et al.~\cite{Berthier04} use a similar way as we for the
calculation of subcritical cluster distributions. To obtain the number density
of clusters of size $s$, they consider the configurational entropy of
the ground and the first two excited states. These are the clusters
with the three lowest perimeters $t$. Compared to our partition
function, which considers all possible configurations (up to the
highest excited state), their summation is incomplete for
$s>7$. Moreover, many of their weights for the excited states are too
low. Since they do not specify the method of counting the accessible
configurations, the origin of the discrepancy is unclear, but it
seems that they miss some configurations. As a consequence,  the
free energies obtained by their method are erroneously non-monotonic
and too high.

The experimentally observed post-deposition dynamics after a potential
pulse,\cite{He01} which result in the continued growth of large clusters while
small clusters already tend to dissolve, are clearly reproduced by our
simulations. Though it is not necessary to explicitly include surface
diffusion under control of the electrochemical potential to establish
lateral mass transport, we find diffusion more likely to be the
dominant mechanism under the experimental conditions. However, in our
simulations, coverages are much higher than in the experiment. While
time scales of adsorption and desorption in the experiment are rather
different, in our model they are necessarily much more similar, since
adsorption and desorption proceed with very similar mechanisms. 

Our model calculations are a first step toward including diffusion into the
microscopic dynamics of a kinetic lattice-gas (or Ising) model and to investigate
its influence on the applicability of the KJMA theory. Still, calculations
over a larger range of fields and a more stringent analysis are
required.\cite{Ramos99} To support our hypotheses about the interplay
between diffusion, nucleation, and spatial
correlation, more microscopic information, like on nucleation events
and the fate of individual clusters, should be obtained from the
simulations. For a more meaningful description of experiments, a
closer reproduction of experimental conditions is desirable.

\begin{acknowledgments}
This work was supported in part by National Science Foundation Grant
No. DMR-0240078, and by Florida State University through its Center
for Materials Research and Technology and School of Computational
Science.
\end{acknowledgments}

\clearpage

\begin{thebibliography}{10}
\bibitem{Footnote1}
For a spin $s_i=\pm 1$ Ising model with $z$ nearest neighbors (here, $z=4$) and
  $\mathcal{N}$ sites, coupling constant $J$ and external magnetic field $H$,
  the mapping between the models is $\mathcal{H}_I=\mathcal{H}_{LG} +
  \frac{1}{2}\mathcal{N}(\mu - \frac{1}{2}\mu_0)$, $c_i=\frac{1}{2}(s_i+1)$, $\phi=4J$,
  $\mu-\mu_0=2H$, and $\mu_0=-2zJ$.
\end{thebibliography}
\end{document}